# An evolving jet from a strongly-magnetised accreting X-ray pulsar

J. van den Eijnden[1], N. Degenaar[1], T. D. Russell[1], R. Wijnands[1], J. C. A. Miller-Jones[2], G. R. Sivakoff[3], J. V. Hernández Santisteban[1]

[1] *Anton Pannekoek Institute for Astronomy, University of Amsterdam, Science Park 904, 1098 XH, Amsterdam, The Netherlands*
[2] *International Centre for Radio Astronomy Research -- Curtin University, GPO Box U1987, Perth, WA 6845, Australia*
[3] *Department of Physics, University of Alberta, CCIS 4-183, Edmonton, AB T6G 2E1, Canada*

**Relativistic jets are observed throughout the Universe, strongly impacting their surrounding environments on all physical scales, from Galactic binary systems[1] to galaxies and galaxy clusters[2]. All types of accreting black holes and neutron stars have been observed to launch jets[3], with one exception: neutron stars with strong (B ≥ $10^{12}$ G) magnetic fields[4,5], creating the theory that their magnetic field strength inhibits jet formation[6]. Radio emission was recently detected in two such objects, with a jet origin being one of several possible explanations in both cases[7,8]. While this provided the first hints that this long-standing idea might need to be reconsidered, definitive observational evidence of jets was still lacking. Here, we report the discovery of an evolving jet launched by a strongly-magnetised neutron star, accreting above the Eddington limit, finally disproving the original theory. The radio luminosity of the jet is two orders of magnitude fainter than seen in other neutron stars at similar X-ray luminosities[9], implying an important role for the neutron star's properties in regulating jet power. This jet detection necessitates fundamental revision of current neutron star jet models and opens up new tests of jet theory for all accreting systems. It also shows that the strong magnetic fields of ultra-luminous X-ray pulsars do not prevent such sources from launching jets.**

On 3 October 2017, the Neil Gehrels Swift Observatory (Swift) detected the outburst of the new X-ray transient Swift J0243.6+6124 (hereafter Sw J0243)[10]. The discovery of 9.86-second pulsations[11] identified this transient as an accreting pulsar: a relatively slowly spinning neutron star with a strong magnetic field (defined as B ≥ $10^{12}$ G)[12], most likely accreting from a high-mass companion Be-star. Throughout its outburst, we observed this source at radio wavelengths over eight epochs with the Karl G. Jansky Very Large Array (VLA), see Figure 1. After an initial non-detection early in the outburst, we detected significant (18.3 sigma) radio emission at 6 GHz close to the X-ray peak (Figure 2), when the neutron star was accreting above the theoretical Eddington limit. The radio luminosity of the system subsequently decayed with the X-ray flux, while the radio spectral index α (where the flux density $S_\nu \propto \nu^\alpha$) gradually evolved throughout the outburst. We did not detect linearly polarised emission during any epoch, with a most stringent upper limit of ~15% during the third observation (see Extended Data for all measurements).

Its radio properties show that Sw J0243 launches an evolving jet. Whenever accreting compact objects launch steady jets, the radio and X-ray luminosity are coupled[9,13] (see Figure 3), indicating a direct relationship between the X-ray emitting accretion flow and the radio-emitting jet. After the initial radio non-detection, we observe such a coupling between the X-ray and radio luminosities of Sw J0243, with the radio luminosity decreasing as the outburst decayed in X-rays. Estimating the correlation index between the 0.5-10 keV X-ray and 6 GHz radio luminosities, we measure $L_R \propto L_X^{0.54 \pm 0.16}$, consistent with both black hole and neutron star X-ray binaries[14] (see Methods).

The radio spectral shape and evolution also argue a jet origin. In radio, jets launched from stellar-mass accretors emit synchrotron radiation with a spectral index that can vary over time, as observed in Sw J0243. The radio spectral index in Sw J0243 started out steep ($\alpha < 0$), only to gradually evolve to a flat spectrum ($\alpha \geq 0$) as observed in canonical steady X-ray binary jets[15]. This systematic evolution during the outburst decay can be interpreted as follows: during the super-Eddington phase, where strong outflows are expected theoretically[16], discrete transient ejecta were launched. When the accretion rate decayed during the remainder of the outburst, the radio — X-ray correlation and the transition towards an inverted spectrum signalled that the radio emission instead arose from a compact, steady jet[15]. Alternatively, a gradual shift of the break frequency where the jet spectrum transitions from optically thin to thick synchrotron radiation could also be responsible for the observed evolution of the radio spectral index. As discussed in the Methods section, alternative physical or emission mechanisms cannot explain the observed combination of spectral index evolution, flux levels, radio — X-ray coupling, and polarisation. We note that both the observed polarization properties and the spectral shape and evolution rule out that coherent radio pulsations are responsible for the radio emission.

Before our radio monitoring campaign of Sw J0243, jets had been confirmed in all types of X-ray binary systems[4,9] except in the strongly-magnetised accreting pulsars, which are the most common X-ray binary type. Multiple large surveys in the 1970s and 80s failed to detect radio emission from these sources[5,17,18] leading to the observational notion that the strong magnetic field prevents the formation of jets. Until recently, searches for radio emission from individual neutron stars with such field strengths also yielded non-detections[6], further strengthening this idea. As a result, strongly-magnetised accreting neutron stars are often disregarded in theoretical studies of neutron star jet formation[19].

Jet formation models developed for accreting neutron stars commonly invoke a magneto-centrifugal launch mechanism[6,20,21], where the jet is launched by field lines anchored in the innermost accretion disk. Such models offered a straightforward theoretical explanation for the prevention of jet formation by strong magnetic fields: the neutron star magnetosphere stops the formation of the inner accretion flow by dominating over the disk pressure[6], therefore preventing the launching of a jet. The first observational results to question this view were the recent radio detections[7,8] of the two strongly-magnetised pulsars Her X-1 and GX 1+4. However, contrary to our Sw J0243 monitoring, both sources were detected at a single frequency during a single epoch, meaning that the origin of the emission remained ambiguous. Given the lack of information on spectral shape, temporal evolution or coupling with the X-ray flux, a jet could neither be excluded nor directly inferred. Moreover, the properties of any putative jets – if present – could not be determined from the limited information available.

Our clear discovery of an evolving jet in Sw J0243 now disproves the long-standing idea that strong magnetic fields prevent the launch of a jet. This directly implies that existing models of jet formation in neutron star X-ray binaries[6,20] need to be revisited. For instance, the jet launching region must be much further out than seen in other classes of jet-forming systems. The presence of X-ray pulsations[11] shows that the magnetosphere dominates the inner accretion flow, channelling the material to the neutron star poles. Conservatively estimated, its minimum size — at the outburst peak, during the first jet detection — is $R_m \gtrsim 320$ gravitational radii (see Methods). Hence, the geometrically thin accretion disk must be truncated much further away from the compact object than typically seen in X-ray binaries with weak magnetic fields (e.g. $\leq 10^9$ G), where the observed jets are thought to be launched close to the accretor[6,20,22]. Moreover, in strongly magnetised pulsars accreting at super-Eddington rates, such as Sw J0243, the magnetosphere might be completely enveloped by accreting material[23]. Such a configuration provides entirely different (geometrical) properties of the inner accretion flow than in other types of X-ray binaries. Interestingly, however, the apparent coupling between the X-ray and radio luminosity during the decay and the spectral index evolution in Sw J0243 are similar to other black hole and neutron star X-ray binaries, but at much higher mass accretion rates. Therefore, it is unclear what similarities there are in the jet formation mechanism and what role the magnetosphere plays.

Despite the phenomenological similarities with jets from stellar-mass black holes and weakly-magnetised neutron stars, the jet in Sw J0243 is orders of magnitude fainter in radio luminosity. This difference is evident in the $L_X$—$L_R$ diagram shown in Figure 3, where Sw J0243 falls two orders of magnitude below other neutron stars accreting at similar super-Eddington X-ray luminosities. Importantly, the only difference with Sw J0243 is that these other neutron stars have a weak magnetic field (e.g. $\leq 10^9$ G) and are spinning faster. Therefore, the difference in radio luminosity might suggest an important role for these fundamental properties of the neutron star in regulating jet power. This role fits with recent theoretical work[21] discussing a neutron star jet model where the jet is powered by the accretor's rotation, as in the Blandford-Znajek-type models for black holes[24], and not launched by field lines in the inner accretion disk, as in the magneto-centrifugal (Blandford-Payne-type) jet models commonly assumed for neutron stars[6,20]. This model, subsequently shown by numerical simulations[25] to be applicable to the super-Eddington accreting regime of Sw J0243, also predicts a two orders of magnitude suppression of jet power for slowly-pulsating, strongly-magnetised accreting pulsars compared to their weakly magnetised, rapidly-spinning counterparts.

Our discovery of a jet in a strongly-magnetised accreting pulsar has two additional major implications. Firstly, it implies that accreting pulsars form a large, hidden class of radio emitters, now accessible for the current generation of observatories with upgraded sensitivities. This unexplored population opens up new avenues to test general predictions of jet theory for all accreting systems. In Blandford-Znajek type models of jet formation, a correlation would be expected between the spin and jet power[21,24]. This straightforward prediction has been difficult to test — estimates of the spin of black holes are challenging, and while pulsations provide an undisputed measurement of neutron star spins, the only neutron stars previously known to launch jets (those with weak magnetic fields) span merely a small range in spin frequency (a factor ~5-6)[26]. Accreting pulsars with strong magnetic fields on the contrary can span over three orders of magnitude in spin, while having similar and well-measured magnetic fields. Now that we have found that strongly-magnetised accreting pulsars can launch jets, future observational campaigns of this source class will probe the predicted relation between spin and jet power.

Secondly, the detection of a jet in Sw J0243 expands the possible types of outflows in ultra-luminous X-ray sources (ULXs), which are binary systems with X-ray luminosities greatly exceeding the Eddington luminosity of a stellar-mass accretor. Super-Eddington winds have previously been observed in both black hole and neutron star ULXs[16], while jets have been inferred in a handful of black hole ULXs through direct detection and the presence of surrounding bubbles[27]. While a number ULXs have been confirmed to be neutron stars through the detection of pulsations, it is now theoretically[28] and observationally[29] realised that such strongly-magnetised ULX pulsars could make up a large fraction of the population. Interestingly, the known ULX pulsars show similar X-ray behaviour to Galactic pulsars accreting from Be stars at super-Eddington rates[30], like Sw J0243. Our detection of a jet in Sw J0243 therefore implies that, in addition to winds, ULX pulsars might also launch jets, unhampered by their strong magnetic fields.

**Acknowledgements** The authors thank the VLA for rapidly accepting and performing the DDT radio observations, and dr. Elena Gallo for providing the black hole sample used in Figure 3. J.v.d.E., N.D. and J.V.H.S. appreciate support from a Netherlands Organisation for Scientific Research (NWO) Vidi grant awarded to N.D. T.D.R. is supported by an NWO Veni grant. R.W. is supported by an NWO Top grant. J.C.A.M.-J. is supported by an Australian Research Council Future Fellowship (FT140101082). G.R.S. acknowledges support from an NSERC Discovery grant. This work made use of data supplied by the UK Swift Science Data Centre at the University of Leicester. The National Radio Astronomy Observatory is a facility of the National Science Foundation operated under cooperative agreement by Associated Universities, Inc. This work has made use of data from the European Space Agency (ESA) mission Gaia

(https://www.cosmos.esa.int/gaia), processed by the Gaia Data Processing and Analysis Consortium (DPAC, https://www.cosmos.esa.int/web/gaia/dpac/consortium). Funding for the DPAC has been provided by national institutions, in particular the institutions participating in the Gaia Multilateral Agreement.

**Author contributions** J.v.d.E. led the two VLA DDT observing proposals, performed the analysis of the Swift data, and wrote the manuscript with comments from all authors. J.v.d.E., N.D. and T.D.R. designed the radio monitoring strategy. T.D.R. and J.v.d.E. jointly analysed the VLA radio data. J.V.H.S. performed the Gaia DR2 distance estimate. All authors made significant contributions to the science case and commented on multiple versions of the manuscript.

**Author information** Reprints and permissions information is available at www.nature.com/reprints. The authors declare no competing financial interests. All correspondence should be addressed to J.v.d.E. (a.j.vandeneijnden@uva.nl)

**Figure 1: Radio and X-ray outburst light curve of Swift J0243.6+6124.**
(a) The radio flux densities at 6 and 22 GHz and the Swift BAT count rate between 15 and 50 keV throughout the outburst. Sw J0243 was not detected during the first radio epoch, marked as downward arrows. (b) The radio spectral index α (where $S_\nu \propto \nu^\alpha$) as a function of time. Error bars are given at the one-sigma level, upper limits at three sigma.

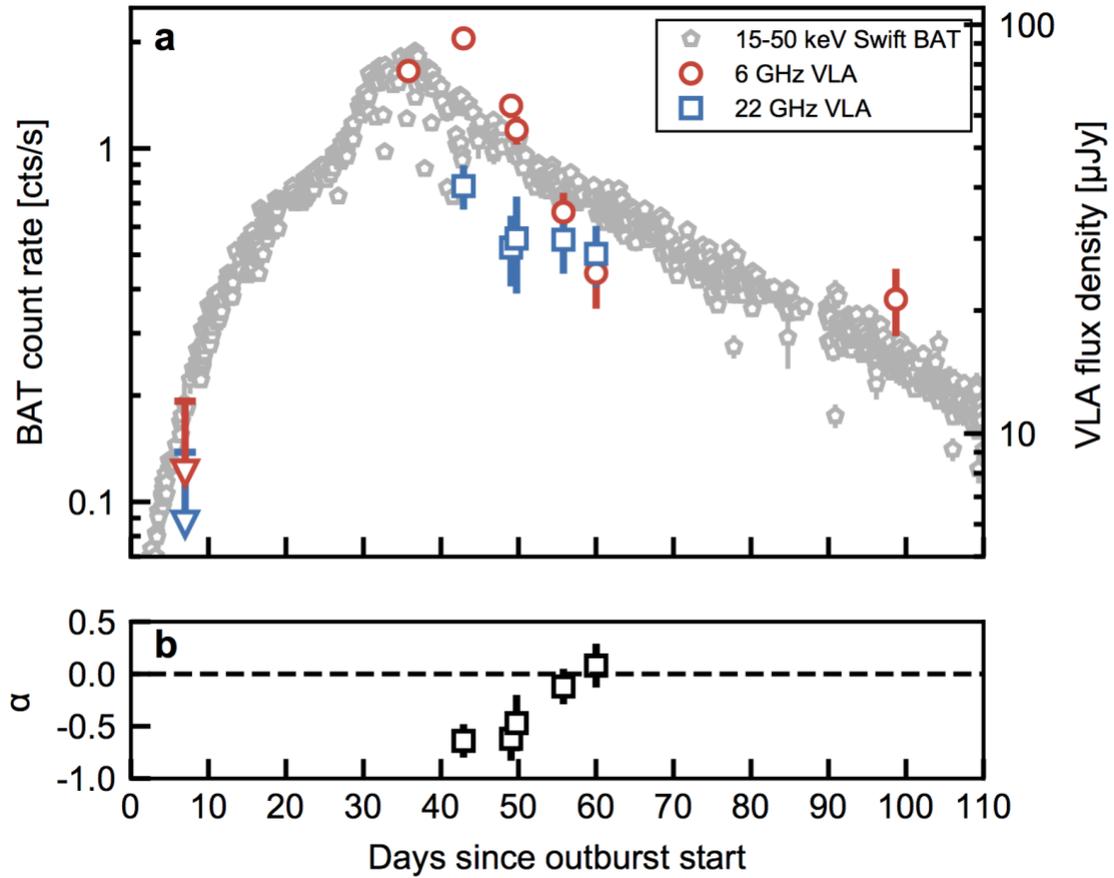

**Figure 2: VLA detection images of Swift J0243.6+6124**
(a) The 6 GHz image of the first VLA observation of Sw J0243. No significant radio emission is observed within the black dashed circle, which indicates the 90% Swift XRT position contour.
(b) The 6 GHz image of Sw J0243 during the second VLA epoch, where the target was first detected. A new, 18.3-sigma significance source is coincident with the Swift XRT position. The synthesised beam is shown in the bottom left corner of both panels.

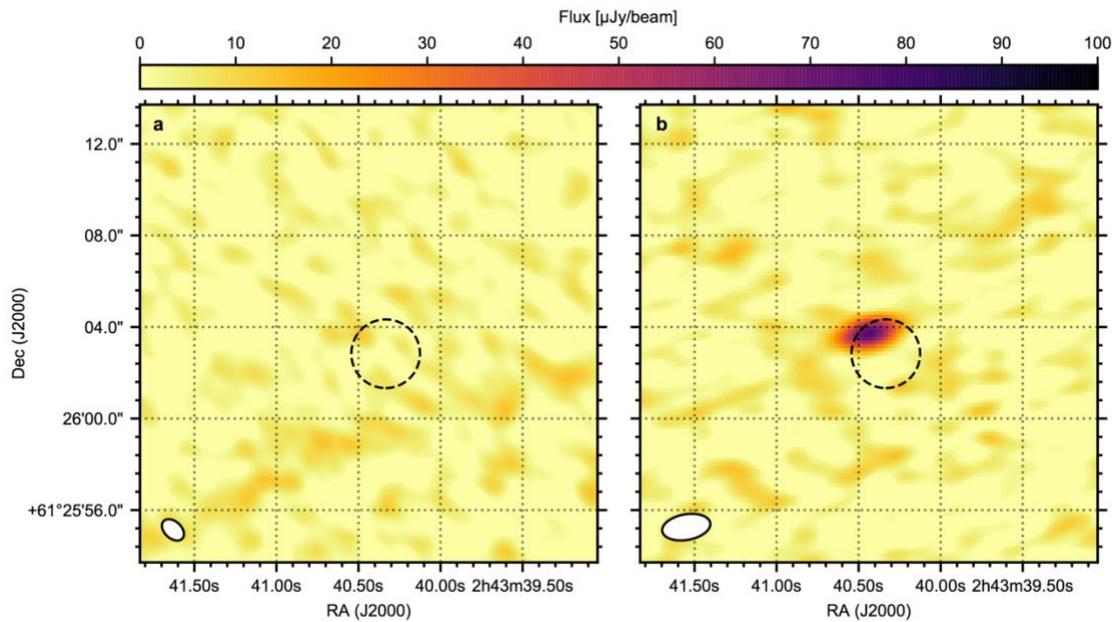

**Figure 3: The radio and X-ray luminosities plane for X-ray binaries.**
The X-ray and radio luminosities of Sw J0243 during the eight epochs, together with a large sample of accreting stellar-mass black holes and neutron stars. The dashed line shows that Eddington X-ray luminosity for a 1.4 $M_{sun}$ neutron star. See the Methods section for details on the sample shown and the estimation of the distance used. Note that for visual clarity we do not plot any non-detections in the sample, or uncertainties of any source in the figure.

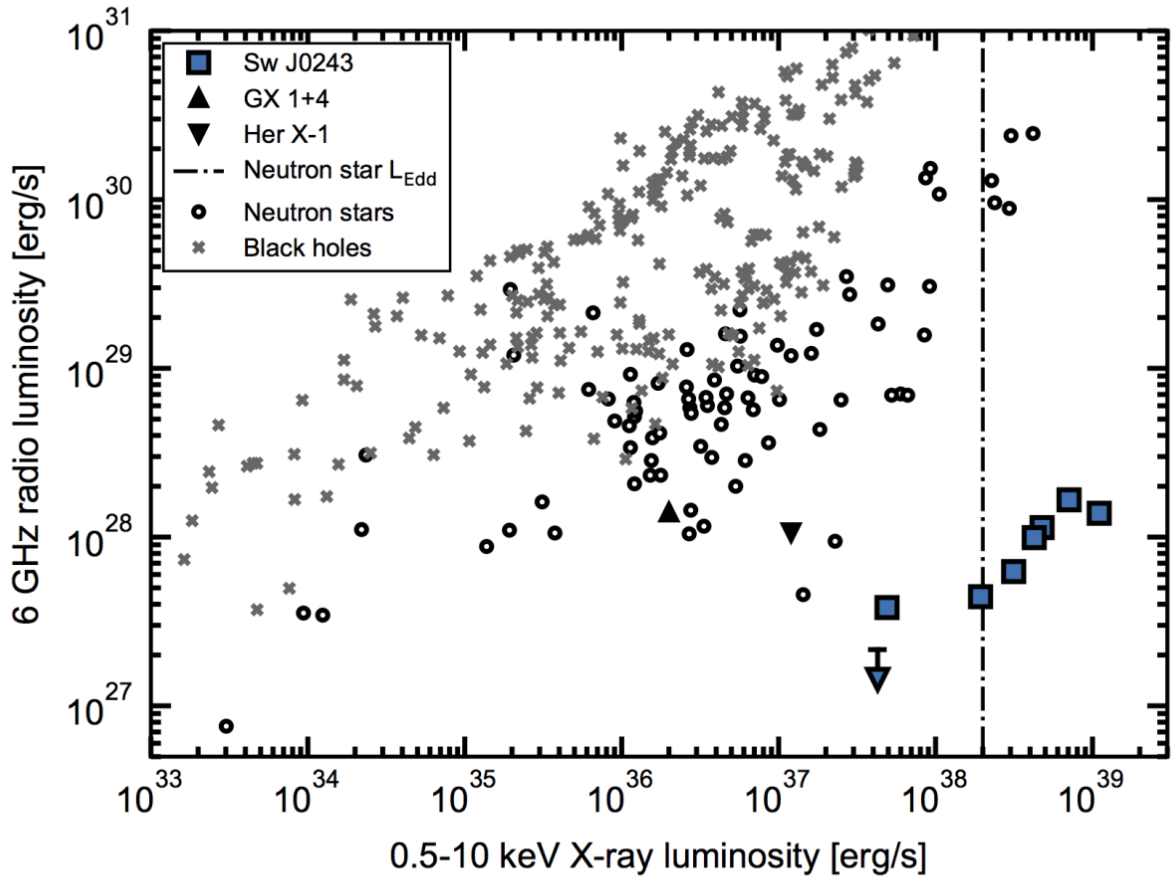

## METHODS

**Radio observations**

We observed Sw J0243 with the VLA over eight epochs between 10 October 2017 and 9 January 2018. The observations were part of two Director's Discretionary Time programs, respectively VLA/17B-406 and VLA/17B-420 for the first two and remaining six epochs. The total observing time was 13 hours. In two observations, we observed the target only at C band, centred at 6 GHz with 4 GHz of bandwidth. In the other six observations, we observed both at C band (with the same setup) and K band, the latter centred at 22 GHz with 8 GHz of bandwidth. Detailed information about each epoch can be found in Extended Data Table 1.

In all epochs, the primary calibrator was J0137+331 (3C48) and the nearby phase calibrator was J0244+6228 (1.04 degrees angular separation to the target). When included in the setup (epochs three to seven), the leakage calibrator was J0319+4130 (3C84). For the target field, the centre was pointed 6" offset in the north direction from the Swift/XRT detection position[10], to prevent possible correlator artefacts at the phase centre from affecting the results. During all observations, the VLA was in its B configuration. More detailed information, such as beam sizes and position angles in each observing band and epoch, are also listed in Extended Data Table 1.

To analyse the observations, we used the Common Astronomy Software Application package[31] (CASA) v4.7.2 to flag, calibrate and image the data. We removed RFI using a combination of automated flagging routines and careful visual inspection. Given the lack of bright radio emission in the target field, we did not self-calibrate. Using the multi-frequency multi-scale CLEAN task with Briggs weighting and a robustness of 1 (to reduce effects of side-lobes of a neighbouring source), we imaged Stokes I at all observed frequencies for all epochs, and Stokes Q and U at 6 GHz for epochs with leakage calibration. We do not image Stokes Q and U at 22 GHz, since we do not detect any linearly polarised emission at 6 GHz. Therefore, no such emission is expected at 22 GHz, and the better RMS at 6 GHz yields tighter upper limits.

Accreting X-ray binaries are expected to be unresolved point sources for the VLA. Therefore, we determined fluxes by fitting an elliptical Gaussian equalling the beam size to the source in the image plane. We measured the RMS of the cleaned image over a region close to the target position. We determined a single flux density in each band and, owing to the faintness of the radio emission, did not divide the C and K band frequency ranges further. A quick check for time variability did not reveal any evidence for significant variability within observations.

The target was not detected in our first observational epoch, with three sigma upper-limits on the flux densities of 12 μJy/beam and 9 μJy/beam at C and K band, respectively. Sw J0243 was detected in all following observations. All flux densities are listed in Extended Data Table 2. The radio position of Sw J0243, measured at 6 GHz from the first detection, is:

RA: 02:43:40.440 ± 0.029s
Dec: +61:26:03.73 ± 0.10"

All positions determined from the radio detections are consistent with the Swift/XRT X-ray position at every epoch. In Extended Figure 1, we show the target field during the initial non-detection and the first detection. The combination of the spatial coincidence between the X-ray and radio position and the coupled X-ray and radio variability, shows that the observed radio emission originates from Sw J0243.

At epochs with both C and K band observations, we calculated the spectral index to investigate the spectral shape. The power law spectral index α (where the flux density $S_\nu \propto \nu^\alpha$) between two frequencies $\nu_1$ and $\nu_2$, with corresponding flux densities $S_1$ and $S_2$, is calculated as:

$$\alpha = \log(S_1/S_2) / \log(\nu_1/\nu_2)$$

To calculate the uncertainty on the spectral index for each individual epoch, we propagate the uncertainties on the measured flux densities and the range in frequencies through a Monte-Carlo simulation: each iteration, two new flux densities are drawn from Gaussian distributions centred on the measured flux densities with standard deviations equalling the measured uncertainties. New frequencies are drawn from uniform distributions over the frequency range of each band. The resulting calculated spectral index is then saved. After $10^6$ iterations, we calculated the spectral index uncertainty as the standard deviation of the simulated spectral indices.

**X-ray flux measurements**
For the study of Sw J0243 in the X-ray luminosity/radio luminosity plane, accurate and precise X-ray fluxes during the radio epochs are required. Three X-ray instruments consistently observed the entire outburst of Sw J0243: the X-ray Telescope (XRT) and Burst Alert Telescope (BAT) instruments aboard Swift[32], and the Monitor of All-sky X-ray Image[33] (MAXI) aboard the International Space Station. The BAT and MAXI are monitoring instruments, while XRT exposures are pointed observations. Both monitoring instruments only provide count rates of observed targets, which can not be converted to a flux straightforwardly without knowing the shape of the X-ray spectrum. The comparison of XRT fluxes and monitoring count rates shows that the broadband X-ray spectral shape evolves during the outburst. This implies that the count rate to flux conversion for the BAT and MAXI is variable as well and therefore makes both instruments inconvenient for accurately estimating the X-ray flux. MAXI is additionally unsuitable, as a visual inspection of the light curve shows several unphysical jumps in count rate pointing towards systematic errors in the monitoring.

The above considerations make the XRT the most reliable to determine the X-ray flux of Sw J0243 during the radio epochs. Five out of the eight radio epochs had quasi-simultaneous XRT coverage (within 2 days). For the remaining three epochs, such XRT observations were not available. However, preliminary flux estimates for all XRT observations, extracted using the Swift-XRT data products generator[34,35], show that Sw J0243 decayed in a steady, log-linear fashion as a function of time. Therefore, for the three radio epochs without close XRT coverage, we estimated the logarithm of the X-ray flux using linear interpolation between the logarithmic fluxes of the preceding and subsequent XRT pointings. Before describing the actual X-ray flux measurements, we stress that the BAT count rate between 15 and 50 keV of Sw J0243 also decayed in a log-linear fashion during our radio monitoring. This implies that the XRT observations, which only provide spectra up to 10 keV, are representative of both the soft and hard X-ray decay of Sw J0243.

We extracted spectra from the radio position of Sw J0243 using the Swift-XRT data products generator[35] and use XSPEC[36] v12.9.0u to fit the data and determine fluxes. All analysed observations were taken in the Window-Timing (WT) mode. We did not use the fluxes provided by the data products generator for our actual measurements; these fluxes are based on a power-law-only model, which is not necessarily accurate for every spectrum. Moreover, the automatic fits are performed between 0.3 and 10 keV, while the WT mode of XRT is subject to calibration uncertainties for moderately to heavily absorbed sources, possibly resulting in poor fits at low energies[37].

We fitted each spectrum with a model containing interstellar absorption, a blackbody component and a power law (TBABS*(BBODYRAD+PO)). As Be/X-ray binaries can have strongly variable local absorption, we did not tie the absorption column between spectra. We assumed Wilms abundances[38] and Verner cross-sections[39], and fitted the spectra in the reliable energy range (0.7 – 10 keV). We then determined unabsorbed fluxes and their uncertainties in the 0.5-10 keV range using CFLUX and the best-fit model. Information on the analysed observations and the fluxes determined in this analysis, including interpolated fluxes, are listed in Extended Data Table 3. The best fit parameters for each spectrum are listed in Extended Data Table 4.

**Gaia distance measurement**
We have used the recent Gaia Data Release 2 [40,41] to obtain an independent measurement of the distance to the system. The measured parallax of Sw J0204 is $\pi = 0.0952 \pm 0.0302$ mas. We followed the standard Bayesian method to infer the distance towards the system[42]. The likelihood function assumes a normal distribution on the GAIA parallaxes and a suggested prior modelled as an exponential decreasing volume density function, with a length scale $L_{sh} = 1.35$ kpc corresponding to the line of sight value[43]. We have taken into account the zero-point from the global astrometric solution $\omega_{zp} = -0.029$ mas[44]. We used a Markov Chain Monte Carlo procedure (as implemented in EMCEE[45]) to sample the posterior distribution of the distance. The marginal posterior distributions are shown in Extended Figure 1. We find a median value of D = 7.3 kpc with 16th and 84th percentiles of 6.1 and 8.9 kpc, respectively. We stress that the posterior distribution is not symmetric and caution should therefore be exercised in using these numbers.

Given the large fractional error of the parallax, the shape of the posterior distribution deviates from a Gaussian distribution and the upper tail is very sensitive to the choice of prior. We investigated the robustness of our distance estimate with different choices of prior, as shown in Extended Figure 1. When using a uniform prior with a maximum distance of 50 kpc, the median of the distribution shifts towards larger distances. However, the lower limit of the distance is greater than 5.0 kpc at ≥ 99% confidence level for the both priors. Therefore, the Gaia measurement shows that the source is located at a distance of at least 5.0 kpc, independent of the prior used. We conservatively adopt this lower limit on the distance in Figure 3.

During the peak of the outburst, around the time of the first radio detection (epoch two), the XRT 0.5 — 10 keV unabsorbed flux is $(3.69 \pm 0.03) \times 10^{-7}$ erg s$^{-1}$ cm$^{-2}$. For the conservative (prior-independent) minimum distance to the source of 5 kpc, this flux corresponds to an X-ray luminosity of $1.1 \times 10^{39}$ (D/5 kpc)$^2$ erg s$^{-1}$. If we apply a bolometric correction, by extrapolating the best model fit to the 0.1 – 100 keV range, we find an even higher luminosity of $1.5 \times 10^{39}$ (D/5 kpc)$^2$ erg s$^{-1}$. The theoretical Eddington luminosity of an accreting neutron star is $2 \times 10^{38}$ erg s$^{-1}$, showing that even for its closest estimated distance, Sw J0243 firmly reached the super-Eddington regime during the outburst.

**Swift BAT light curve**
To show the long-term X-ray evolution of Sw J0243, we display the Swift BAT light curve in Figure 1. However, for clarity, we show a cleaned version of this light curve: due to the extreme count rates of the source, the measured count rate sometimes dropped down by an order of magnitude in individual exposures. The BAT team ascribes these dropouts to software issues and not to intrinsic variability in Sw J0243[46]. Therefore, we masked these anomalously low points in the light curve, which occur between 19 and 60 days into the outburst, around the times with the highest count rates. In this time interval the actual rates exceeded 0.7 counts per second, so we removed all exposure with lower count rates. We stress that this cleaning procedure is for visual purposes only and does not affect our actual measurements or conclusions.

**Estimating the magnetospheric radius**
The magnetospheric radius is defined as the radius where the pressure of the magnetosphere and accreting material are equal. Therefore, this radius will depend on the strength of the magnetic field B and the rate of accretion. The latter can be estimated from the bolometric flux F, the distance D (together providing the X-ray luminosity), the accretion efficiency ή (converting the mass accretion rate to luminosity) and an anisotropy correction factor *f* to account for the anisotropy of the emitted X-rays. Finally, the type of accretion (i.e. wind or disk) has to be taken into account through a geometrical correction factor *k*. For standard neutron star parameters — a mass of 1.4 Solar mass and a radius of 10 km — the magnetospheric radius $R_m$ (in gravitational radii $R_g = GM/c^2$) can be estimated from the above parameters as[47,48,49]:

$$R_m = k \, (B / 1.2 \times 10^5 \text{ G})^{4/7} \, (f / ή)^{-4/14} \, (F / 10^{-9} \text{ erg s}^{-1} \text{ cm}^{-2})^{-4/14} \, (D / 5 \text{ kpc})^{-4/7} \, R_g$$

While not all parameters are known precisely, we can use this equation to estimate a minimum size of the magnetosphere during the outburst. The maximum unabsorbed, bolometric X-ray flux observed by Swift/XRT around a radio epoch, which will give the smallest magnetospheric radius, is $4.9 \times 10^{-7}$ erg s$^{-1}$ cm$^{-2}$ (but see below). Gaia measurements with an exponential prior imply a median distance estimate of 7.3 kpc, which we will adopt for this calculation – this provides a more conservative lower limit on $R_m$ than using the 5 kpc minimum distance. The minimum value of *k* is 0.5, as appropriate for disk accretion[50]. The accretion efficiency is typically assumed to be 0.1, while the anisotropy correction is close to unity[48]. Finally, the magnetic field is not measured directly but can be determined from the X-ray pulsations[12] to exceed $10^{12}$ G. Combining these numbers yields $R_m \gtrsim 320 \, R_g$ (i.e. ~670 km).

Following typical assumptions, we have used the bolometric X-ray flux, combined with an efficiency of 10%, to probe the mass accretion rate that balances the magnetospheric pressure. However, a non-negligible fraction of the X-ray flux might be emitted from the neutron star surface with higher efficiency, which would imply that this approach might overestimate the mass accretion rate. On the other hand, outflows from the neutron star or disc would cause the flux-derived mass accretion rate to be underestimated. Given these contradictory possibilities, we do not correct for either of these processes: correcting for the former will lead to a larger magnetospheric radius, which is already consistent with our approach of calculating a lower limit; correcting for the latter will lead to a lower radius, but this correct is small given the weak scaling between mass accretion rate and magnetospheric radius (-2/7). Thus, correcting for either case does not affect our conclusions.

**Radio — X-ray correlation sample**

For the discussion of the radio — X-ray correlation, shown in Figure 3, we use a comprehensive sample of hard state Atoll neutron star sources and hard state black holes, collected from the large body of observational studies of X-ray binaries performed over the past decades. This sample is freely available online[i] and was originally collected for a different study focussing on the radio — X-ray plane of accreting neutron stars[14]. To this sample, we add the Z-sources[9], two jet-quenched accreting neutron stars[51,52] and the accreting pulsars[7,8] GX 1+4 and Her X-1. We add these sources as interesting comparisons with Sw J0243: the Z-sources have similar X-ray luminosities, the jet-quenched neutron stars have similar radio luminosities, and the accreting pulsars have similar physical characteristics. Note that, as discussed extensively in the next section, it remains unclear whether the radio emission from these two accreting pulsars originates from a jet.

The radio luminosities in the full sample are collected at 5 GHz, while we measured the 6 GHz radio luminosity of Sw J0243 in our monitoring. Hence, we transformed the 5 GHz sample luminosities to 6 GHz by assuming a flat spectrum, which amounts to multiplying all luminosities in the sample by 6/5. Assuming a flat radio spectrum will not be accurate for each observation. For instance, a clear effect of the radio spectral shape on the position of black hole systems on the radio — X-ray plane has recently been demonstrated[53]. However, making this simplifying assumption is valid since we show the large sample only for a broad qualitative comparison between Sw J0243 and other types of sources. Our conclusions — that Sw J0243 shows an apparent coupling between in- and outflow as well and is 2 orders of magnitude fainter than the Z-sources — are not affected by assuming the flat radio spectrum.

Finally, we note that we plot the 0.5 — 10 keV X-ray luminosity of Sw J0243 to be consistent with the full sample. Before Sw J0243, no radio emission confirmed to be from a jet had been detected from any confirmed high-mass X-ray binary system containing a neutron star. Therefore, all neutron stars in the sample reside in low-mass X-ray binaries. While the 0.5 — 10 keV X-ray luminosity does not necessarily probe the same components of the accretion flow in low and high-mass X-ray binaries, we plot this energy range to remain consistent between all sources and with the existing literature.

**Measuring the X-ray – radio correlation index**

We measured the correlation index from the 0.5-10 keV X-ray and 6 GHz radio luminosities in epochs 2 to 8 (i.e. those with radio detections). We fit the following function to these seven data points:

$L_R = L_{R,\text{ref}} (L_X/L_{X,\text{ref}})^\beta$

---

[i] https://jakobvdeijnden.wordpress.com/radioxray/

where $L_{X,ref}$ is the average X-ray flux of all epochs and $L_{R,ref}$ and $\beta$ are free parameters. We find $\beta = 0.54 \pm 0.16$, consistent with the indices for both the black hole and weakly-magnetised neutron star X-ray binaries[14].

It is important to treat this measurement with caution. Our monitoring of Sw J0243 spans a factor of approximately twenty in X-ray luminosity and five in radio luminosity during the outburst. However, to accurately measure the coupling index between the radio and X-ray luminosities, detailed monitoring over at least two orders of magnitude in X-ray luminosity is strongly recommended[54]. Therefore, while consistent with other X-ray binaries, the exact value is not necessarily representative of the entire outburst or accreting pulsars in general.

From the $L_X - L_R$ diagram, it is clear that without including the lowest X-ray luminosity radio detection, the correlation index would be steeper. While we cannot draw conclusions based on a single data point, this might reflect changes in the jet properties as the source become sub-Eddington and the accretion flow geometry changes.

**Alternative interpretations**
Here we will briefly discuss a number of alternative interpretations for the observed radio properties of Sw J0243. As mentioned in the main paper, none of these alternative explanations can account for the observed combination of radio — X-ray coupling, flux levels, spectral index evolution and polarisation properties.

Firstly, the stellar wind in a high-mass X-ray binary system can emit in radio. Through a combination of optically thick and thin free-free processes, the radio spectrum of such a wind could be flat[55] (i.e. $\alpha = 0$), as we observe in later epochs (see Figure 1 and Extended Data Table 1). However, the systematic evolution seen in the spectral index, similar to that in low-mass X-ray binaries, is not expected for a stellar wind. The same goes for the clear coupling between radio and X-ray flux.

We can also consider the flux levels expected from a stellar wind. The typical flux $S_\nu$ of a stellar wind can be estimated[7,8,55,56] for a given mass accretion rate $\dot{M}$, velocity v, distance D and observing frequency $\nu$:

$$S_\nu = 7.26 \, (\nu / 10 \text{ GHz})^{0.6} \, (\dot{M} / 10^{-6} \text{ solar mass yr}^{-1})^{4/3} \, (v / 100 \text{ km s}^{-1})^{-4/3} \, (D / 1 \text{ kpc})^{-2} \text{ mJy},$$

where we ignore the electron temperature due to its negligible effect on the predicted flux and assume a hydrogen wind (which yields the highest predicted flux). Conservatively assuming the escape velocity of a typical Be-star as a minimum for the wind velocity, and using the lower limit on the distance of 5 kpc (yielding the highest flux density) at a frequency of 6 GHz, we find that the mass-loss rate in the wind needs to exceed $10^{-5}$ solar masses per year to account for the observed flux levels around the outburst peak. Such rates are only associated with Wolf-Rayet stars and are highly unlikely for a Be-star, which are more likely to lose mass at maximally $10^{-9}$ solar masses per year[57]. At rates of $10^{-9}$ solar masses per year, the wind flux would not be expected to exceed 0.01 µJy, orders of magnitude below our radio detections.

Alternatively, accreting neutron stars could launch an outflow through the propeller mechanism. If the rotational velocity of the accreting material is lower than the neutron star spin at the magnetospheric radius, the material can be expelled in a propeller outflow[58]. However, given the magnetic field strength and spin of Sw J0243, such an outflow is not expected at the high mass-accretion rates present when we detect radio emission[59], as the magnetospheric radius is then pushed far inside the co-rotation radius. Instead, the propeller regime and its associated outflows are expected more typically below $10^{36}$ erg s$^{-1}$ for Be/X-ray binary systems[60], which is over two orders of magnitude below the super-Eddington X-ray luminosities of Sw J0243.

Radio pulsations at the neutron star spin frequency also cannot be the origin of the observed emission. While Sw J0243 is too faint to explicitly search for pulsations at the known spin, the spectral shape and evolution rule out this origin: radio pulsations have a steep ($\alpha \cong -1.4$) spectrum that does not evolve[61], contrary to the different and evolving spectral shape observed in Sw J0243.

Coherent emission of any form is ruled out due to the lack of observed circular polarization in Sw J0243 at any epoch.

Finally, shocks between the accreting material and the magnetosphere could give rise to radio emission. However, while the luminosity of this mechanism could naively be expected to scale with the accretion rate and thus X-ray luminosity, we do not necessarily expect the shock spectrum to evolve as observed: the regular evolution of the spectrum from optically thin to thick, coupled to the decaying X-rays, implies that the same mechanism is responsible for all emission. The spectral shape towards the end of our radio monitoring, i.e. flat, is inconsistent with the optically thin spectrum expected for the shocked emission.

Therefore, none of these alternative mechanisms can account for our radio observations of Sw J0243. The observed radio properties directly point towards a jet origin (as argued in the main paper). Combined, the exclusion of alternatives and direct implication of a jet origin make Sw J0243 a completely distinct case from Her X-1 and GX 1+4. While those strongly-magnetised accreting neutron stars were recently detected in radio, these single-frequency/single-epoch detection could not directly imply a jet origin[7,8]. Several of the alternative mechanisms discussed above could also not be excluded. Therefore, while inspiring our multi-band monitoring campaign of Sw J0243, those detections could neither convincingly prove that there were jets in strongly magnetised neutron stars (thus disproving the existing theory) nor provide details on the properties of such jets.

**Code availability** the code used to estimate the distance from the Gaia DR2 measurements is available at https://github.com/Alymantara/Sw_J0243. All data analysis software is publicly available for download (CASA: https://casa.nrao.edu, HEASOFT: https://heasarc.nasa.gov/lheasoft/). This research made use of Astropy, a community-developed core Python package for Astronomy[62], available at https://www.astropy.org.

**Data availability** the VLA observations analysed in the work will become publicly available in the NRAO Science Data Archive (https://archive.nrao.edu/archive/advquery.jsp) on 8 November 2018 (first two epochs) and 20 February 2019 (remaining epochs), using respectively Project

Code 17B-406 and 17B-420. However, prior access to the VLA observations will be granted by the corresponding author upon reasonable request. All Swift X-ray data is accessible in the HEASARC data archive. The radio — X-ray correlation data sample is available online at https://github.com/jvandeneijnden/XRB-Lx-Lr-Sample.

**Extended Data Table 1: Overview of VLA radio observations of Sw J0243**
For each radio epoch, we list the start and end time of the target observations in UTC (i.e. not including initial setup and calibration), the observing frequencies, whether we observed a leakage calibrator, and the beam size and position angle (degrees east of north) at each frequency. The 6 GHz observations were performed with 4 GHz of bandwidth; the 22 GHz observations with 8 GHz of bandwidth.

| Radio Epoch | Start (UTC) | End (UTC) | Observing frequencies | Leakage Calibrator | Beam size (position angle) |
|---|---|---|---|---|---|
| 1 | 2017-10-10 05:37:00 | 2017-10-10 06:09:50 | 6 GHz | No | 1.12" x 0.74" (47.4 deg) |
|   |   |   | 22 GHz | No | 0.30" x 0.20" (55.3 deg) |
| 2 | 2017-11-08 00:46:52 | 2017-11-08 01:29:42 | 6 GHz | No | 2.13" x 1.12" (-80.0 deg) |
| 3 | 2017-11-15 03:21:21 | 2017-11-15 04:46:36 | 6 GHz | Yes | 1.32" x 0.99" (35.3 deg) |
|   |   |   | 22 GHz | No | 0.40" x 0.29" (52.2 deg) |
| 4 | 2017-11-21 06:27:33 | 2017-11-21 07:52:46 | 6 GHz | Yes | 1.42" x 0.97" (-30.9 deg) |
|   |   |   | 22 GHz | No | 0.34" x 0.28" (-17.1 deg) |
| 5 | 2017-11-22 23:29:24 | 2017-11-23 00:54:36 | 6 GHz | Yes | 1.89" x 1.06" (-87.6 deg) |
|   |   |   | 22 GHz | No | 0.57" x 0.33" (-74.1 deg) |
| 6 | 2017-11-28 00:26:51 | 2017-11-28 01:58:04 | 6 GHz | Yes | 1.54" x 1.02" (70.7 deg) |
|   |   |   | 22 GHz | No | 0.45" x 0.27" (82.3 deg) |
| 7 | 2017-12-02 05:47:30 | 2017-12-02 07:20:14 | 6 GHz | Yes | 1.33" x 1.03" (-31.2 deg) |
|   |   |   | 22 GHz | No | 0.35" x 0.28" (-12.8 deg) |
| 8 | 2018-01-09 22:14:52 | 2018-01-09 22:57:44 | 6 GHz | No | 1.61" x 1.10" (76.9 deg) |

**Extended Data Table 2: VLA radio flux density, polarisation and position measurements**
For each radio epoch and observing frequency, we show the observed flux densities or 3-sigma upper limits in case of non-detection, the spectral index when both 6 and 22 GHz observation were carried out, the most stringent upper limit on linear polarisation per epoch if available, and the 6 GHz position per epoch. All uncertainties are at 1-sigma, while upper limits are quoted at three sigma. The errors on the position are calculated by taking the maximum of the synthesised beam size divided by the signal-to-noise of the source detection and 10% of synthesised beam size, following VLA guidelines.

| Radio Epoch | Observing frequency | Flux density [µJy] | Spectral index $\alpha$ | Linear polarisation | 6 GHz position |
|---|---|---|---|---|---|
| 1 | 6 GHz | < 12.0 | - | - | - |
|   | 22 GHz | < 9.0 |   |   |   |
| 2 | 6 GHz | 77.1 ± 4.2 | - | - | RA: 02:43:40.440 ± 0.029s<br>Dec: +61:26:03.73 ± 0.10" |
| 3 | 6 GHz | 92.6 ± 3.8 | -0.64 ± 0.16 | < 17% | RA: 02:43:40.425 ± 0.022s<br>Dec: +61:26:03.73 ± 0.18" |
|   | 22 GHz | 40.3 ± 5.0 |   |   |   |
| 4 | 6 GHz | 63.4 ± 4.3 | -0.62 ± 0.21 | < 27% | RA: 02:43:40.419 ± 0.015s<br>Dec: +61:26:03.80 ± 0.13" |
|   | 22 GHz | 28.5 ± 5.6 |   |   |   |
| 5 | 6 GHz | 55.3 ± 4.4 | -0.47 ± 0.27 | < 34% | RA: 02:43:40.430 ± 0.026s<br>Dec: +61:26:03.74 ± 0.10" |
|   | 22 GHz | 30.0 ± 8.0 |   |   |   |
| 6 | 6 GHz | 34.8 ± 4.0 | -0.12 ± 0.17 | < 47% | RA: 02:43:440 ± 0.024s<br>Dec: +61:26:03.65 ± 0.13" |
|   | 22 GHz | 29.8 ± 5.2 |   |   |   |
| 7 | 6 GHz | 24.7 ± 4.5 | 0.08 ± 0.21 | < 75% | RA: 02:43:40.419 ± 0.028s<br>Dec: +61:26:03.69 ± 0.23" |
|   | 22 GHz | 27.5 ± 4.7 |   |   |   |
| 8 | 6 GHz | 21.3 ± 4.0 | - | - | RA: 02:43:40.432 ± 0.042s<br>Dec: +61:26:03.97 ± 0.21" |

**Extended Data Table 3: Swift XRT flux measurements**
For each radio epoch, we list the Swift XRT observations used to determine the unabsorbed X-ray flux. When two ObsIds are listed, the X-ray flux estimate for that radio epoch was determined through log-linear interpolation between the two shown observations. Note that three leading zeros have been removed from all ObsIds. All errors are quoted at 1-sigma.

| Radio Epoch | Swift XRT ObsId(s) | Start date(s) | Unabsorbed flux [$10^{-8}$ erg s$^{-1}$ cm$^{-2}$] | Interpolated flux [$10^{-8}$ erg s$^{-1}$ cm$^{-2}$] |
|---|---|---|---|---|
| 1 | 10336007 | 2017-10-10 | 1.43 ± 0.01 | n/a |
| 2 | 10336022 | 2017-11-09 | 36.9 ± 0.25 | n/a |
| 3 | 10336025 | 2017-11-15 | 23.7 ± 0.16 | n/a |
| 4 | 10336025 | 2017-11-15 | 23.7 ± 0.16 | 16.0 ± 0.11 |
|   | 10336031 | 2017-11-27 | 10.5 ± 0.07 |  |
| 5 | 10336025 | 2017-11-15 | 23.7 ± 0.16 | 14.2 ± 0.10 |
|   | 10336031 | 2017-11-27 | 10.5 ± 0.07 |  |
| 6 | 10336031 | 2017-11-27 | 10.5 ± 0.07 | n/a |
| 7 | 10336033 | 2017-12-01 | 6.47 ± 0.04 | n/a |
| 8 | 10467007 | 2018-01-02 | 2.04 ± 0.01 | 1.64 ± 0.01 |
|   | 10467008 | 2018-01-13 | 1.47 ± 0.01 |  |

**Extended Data Table 4: Swift XRT spectral fit parameters**

For each analysed Swift XRT observation, we list the best fit spectral parameters for a TBABS*(BBODYRAD+POWERLAW) model in XSPEC. As in Be/X-ray binaries local absorption can contribute to the total absorption column, we do not tie $N_H$ between observations. In observation 10467008, the inclusion of a blackbody spectral component was not statistically required. All errors are quoted at 1-sigma.

| Swift XRT ObsId | $N_H$ [$10^{22}$ cm$^{-2}$] | $T_{BB}$ [keV] | $N_{BB}$ | $\Gamma$ | $N_{PO}$ [phot/keV/cm$^2$/s] | $\chi^2 / \nu$ |
|---|---|---|---|---|---|---|
| 10336007 | 1.60 ± 0.10 | 1.94 ± 0.06 | 43 ± 8 | 1.96 ± 0.16 | 1.85 ± 0.26 | 1028.8 / 865 |
| 10336022 | 1.19 ± 0.08 | 2.08 ± 0.06 | 1036 ± 162 | 1.84 ± 0.15 | 39.2 ± 4.6 | 814.3 / 877 |
| 10336025 | 1.28 ± 0.08 | 1.96 ± 0.05 | 810 ± 104 | 1.99 ± 0.14 | 28.7 ± 3.2 | 967.5 / 874 |
| 10336031 | 1.57 ± 0.09 | 1.96 ± 0.04 | 387 ± 46 | 2.13 ± 0.16 | 13.3 ± 1.7 | 944.6 / 873 |
| 10336033 | 1.46 ± 0.07 | 1.96 ± 0.04 | 212 ± 24 | 1.77 ± 0.13 | 6.56 ± 0.65 | 1007.4 / 889 |
| 10467007 | 1.47 ± 0.08 | 1.75 ± 0.06 | 80 ± 10 | 1.81 ± 0.13 | 2.42 ± 0.25 | 930.4 / 876 |
| 10467008 | 1.42 ± 0.04 | - | - | 1.31 ± 0.02 | 1.48 ± 0.04 | 1058.2 / 870 |

**Extended Data Figure 1: Marginal posterior distributions for the distance to Sw J0243.**
We show the distribution for an exponential and a uniform prior. The median value (50th percentile) of the distribution for the exponential prior is shown as the dot-dashed line.

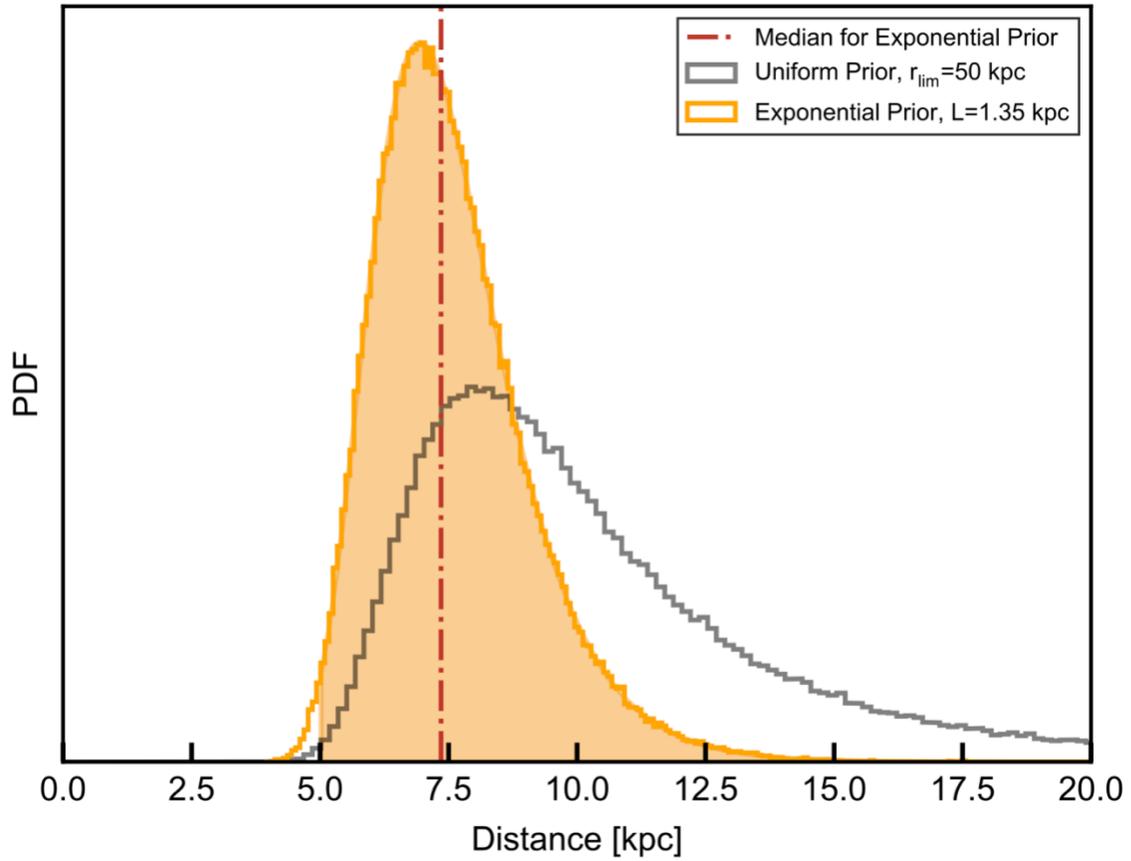